\pdfoutput=1

\documentclass[11pt]{article}

\usepackage{acl}
\usepackage{lipsum}
\usepackage{booktabs}
\usepackage{url}
\usepackage{multirow}
\usepackage{enumitem}
\usepackage{amsmath,graphicx}
\usepackage{makecell}
\usepackage{tabularx}

\usepackage{inconsolata}
\usepackage{bbding}
\usepackage{pifont}
\usepackage{tablefootnote}

\newcommand{\tabincell}[2]{\begin{tabular}{@{}#1@{}}#2\end{tabular}}

\usepackage{times}
\usepackage{latexsym}

\usepackage[T1]{fontenc}

\usepackage[utf8]{inputenc}

\usepackage{microtype}

\usepackage{inconsolata}

\title{Speech-to-Speech Translation with Discrete-Unit-Based Style Transfer}

\author{Yongqi Wang, Jionghao Bai, Rongjie Huang, Ruiqi Li, Zhiqing Hong, Zhou Zhao \\
Zhejiang University \\
\texttt{cyanbox@zju.edu.cn} 
}

\begin{document}
\maketitle
\begin{abstract}
Direct speech-to-speech translation (S2ST) with discrete self-supervised representations has achieved remarkable accuracy, but is unable to preserve the speaker timbre of the source speech. Meanwhile, the scarcity of high-quality speaker-parallel data poses a challenge for learning style transfer during translation. We design an S2ST pipeline with style-transfer capability on the basis of discrete self-supervised speech representations and codec units. The acoustic language model we introduce for style transfer leverages self-supervised in-context learning, acquiring style transfer ability without relying on any speaker-parallel data, thereby overcoming data scarcity. By using extensive training data, our model achieves zero-shot cross-lingual style transfer on previously unseen source languages. Experiments show that our model generates translated speeches with high fidelity and speaker similarity. \footnote{ Audio samples are available at \url{http://stylelm.github.io/} }
\end{abstract}

\section{Introduction}

\begin{figure*}[htb]
\includegraphics[width=\textwidth]{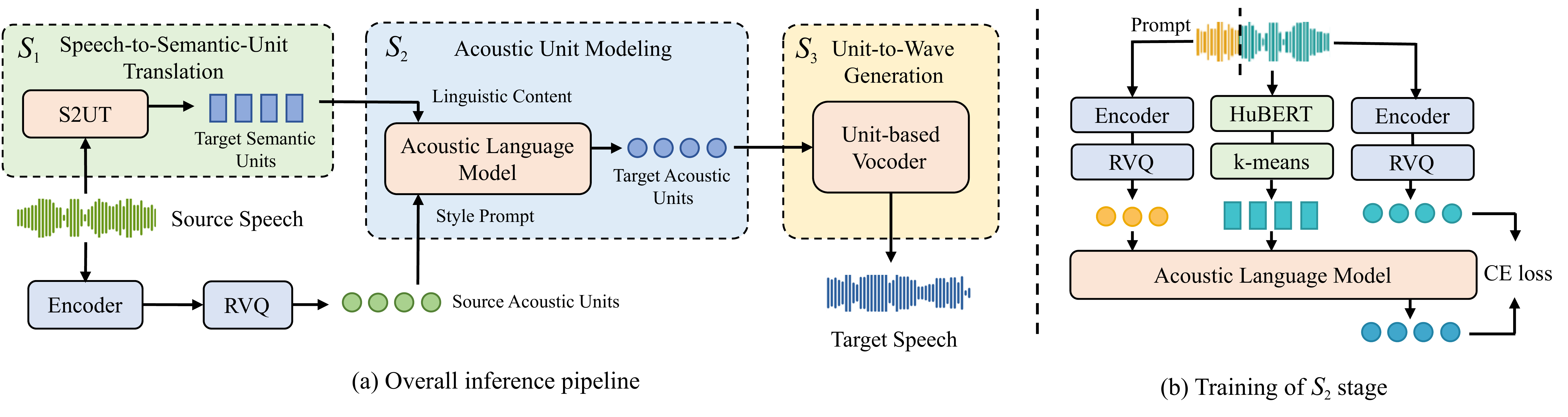}
\caption{We propose an S2ST approach with style transfer based on discrete representations from a self-supervised speech model and a neural codec. Figure (a) shows the inference pipeline of our method; figure (b) illustrates the self-supervised training process of the acoustic language model of $S_2$. }
\label{fig:main}
\end{figure*}

Speech-to-speech translation (S2ST) aims to translate spoken utterances from one language to another, which can bring immense convenience to international communication. Compared to conventional cascaded systems comprising ASR, text translation, and TTS models \cite{lavie1997janus,nakamura2006atr,wahlster2013verbmobil}, direct S2ST models without intermediate text generation have a more concise pipeline with less computation cost and error propagation, and also facilitates application to unwritten languages, and thus spark widespread interest in the community.

Mainstream approaches of direct S2ST \cite{lee2021direct, lee2021textless, huang2022transpeech, popuri2022enhanced} utilize discrete speech representation from self-supervised models (such as HuBERT \cite{hsu2021hubert}) as prediction target, and then use them to reconstruct the waveform. Such representation eliminates speaker identity and prosody of the speeches and retains only semantic contents, which simplifies the target distribution and makes the translation less challenging. However, it also has the drawback of losing the style information of the source speech. Extra voice conversion systems are needed if users want to keep the source speaker timbre, which may cause degradation in audio quality.

Some works propose direct S2ST with style transfer \cite{jia2021translatotron, song2023styles2st}. These methods depend on paired data that source and target speech share the same speakers. However, such data from the real world is extremely scarce as it requires a large number of multilingual speakers, while simulated data from multilingual TTS systems suffers from less diversity and extra data collection costs. Recent large-scale S2ST models \cite{rubenstein2023audiopalm, barrault2023seamless} have also incorporated the capability of style transfer, yet their sub-modules are highly coupled and are difficult to apply to other S2ST models.
 
Inspired by recent progress in spoken language models \cite{borsos2023audiolm, wang2023neural}, we propose a novel approach for direct S2ST with the ability of cross-lingual style transfer, and does not rely on any speaker-parallel data. We utilize two types of discrete representations, namely semantic and acoustic units, from a self-supervised speech model and a neural codec, separately. Our method encompasses three stages: 1) speech-to-semantic-unit translation, which translates source speech to target semantic units; 2) acoustic unit modeling, which generates target acoustic units from translated semantic units using style information in the source speech; and 3) unit-to-wave generation, which reconstructs high-fidelity translated speech from the acoustic units. The modules of the three stages are trained independently and decoupled from each other, allowing our framework to be applied to various existing speech-to-unit translation models.

For the acoustic unit modeling stage, we introduce an acoustic language model. It employs a self-supervised training approach and learns style transfer through in-context learning, which relies on no speaker-parallel data, and thus addresses the issue of data scarcity. By utilizing extensive training data, our model achieves zero-shot cross-lingual style transfer with source languages not included in the training. Experiments show that our model generates results with superior audio quality and style similarity while maintaining accurate content to a good extent.

Our contributions can be summarized as follows:

\begin{itemize}
    \item We propose an S2ST approach with cross-lingual style transfer capability, even on previously unseen source languages.
    
    \item By employing self-supervised training, 
    our model does not rely on any speaker-parallel data, thus addressing the issue of data scarcity.

    \item The decoupling nature of the sub-modules enables our framework to be adopted by various existing speech-to-unit translation models.
    
    \item Experiments show that our method generates translated speeches with high quality and style similarity.
\end{itemize}

\section{Method}
\label{sec:method}

The overall inference pipeline of our method is illustrated in Fig.\ref{fig:main} (a). Our method comprises three consecutive stages, utilizing two distinct types of discrete units: 1) speech-to-semantic-unit translation stage $S_1$, which converts source audio into semantic units of the translated speech; 2) acoustic unit modeling stage $S_2$, generating target acoustic units conditioned on the semantic output from the preceding stage and the acoustic units of the source speech as style prompt; 3) unit-to-wave generation stage $S_3$, producing translated speech that maintains consistent style with the source. We provide details about these two types of units and the three stages in the following subsections.

\subsection{Semantic and Acoustic Units}

Discrete HuBERT \cite{hsu2021hubert} units obtained from the clustering of self-supervised speech representations are shown \cite{lee2021textless, huang2022transpeech} to be effective in providing semantic content information and are widely adopted in S2ST as prediction target \cite{lee2021direct, lee2021textless, huang2022transpeech, popuri2022enhanced}. HuBERT encodes the target speech into continuous representations with a frame length of 20 ms, and these representations are then discretized with the k-means algorithm to get the semantic units.

On the other hand, audio codec models with encoder-decoder architecture such as SoundStream \cite{zeghidour2021soundstream} have recently shown outstanding performance in learning acoustic information. Such a codec model can produce discrete representations (i.e. the acoustic units) of audio by employing a convolutional encoder followed by a residual vector quantizer. These representations contain detailed acoustic information and can be used to reconstruct waveforms with the corresponding decoder or an additional vocoder.

\subsection{Speech-to-Semantic-Unit Translation}

The speech-to-semantic-unit translation stage generates translated semantic units conditioned on source speech input, achieving translation of linguistic content. Various models \cite{lee2021direct,  huang2022transpeech, popuri2022enhanced} have been proposed for this procedure. These models share a common basic architecture of a convolutional speech encoder followed by an encoder-decoder architecture based on a transformer \cite{vaswani2017attention} or conformer \cite{gulati2020conformer}. Due to the decoupling nature of the sub-modules of the three stages, we have the flexibility to adopt different S2UT models in this stage, and we attempted two of them in our experiments (See Section \ref{sec:setup}).

\subsection{Acoustic Unit Modeling}

The acoustic unit modeling stage $S_2$ generates translated acoustic units from semantic tokens and style prompts. The core component of $S_2$ is an acoustic language model, which is basically a decoder-only transformer. Specifically, we adopt UniAudio \cite{yang2023uniaudio} as the acoustic language model, which is proven to be an effective autoregressive audio generation model. Details of the model architecture are provided in Appendix \ref{sec:uniaudio}. The model takes a prefix sequence formed by concatenating acoustic unit sequence $\mathbf{a_p}$, which serves as a style prompt, and the target semantic sequence $\mathbf{s}$, and generates the target acoustic sequence $\mathbf{a}$ with autoregressive sampling. This procedure can be formulated as
\begin{equation}
    \footnotesize
    p\left(\mathbf{a} \mid \mathbf{a}_{\mathbf{p}}, \mathbf{s} ; \theta_{A R}\right)=\prod_{t=1}^T \prod_{c=1}^C p\left(\mathbf{a}_t^c \mid \mathbf{a}_{<t}, \mathbf{a}_{t}^{<c}, \mathbf{a}_{\mathbf{p}}, \mathbf{s} ; \theta_{A R}\right)
\end{equation}
The entire sequence is in the format of $[\mathbf{a_p | s | a}]$, with a separator token between each pair of adjacent parts. 3 codebooks are used for $\mathbf{a_p}$ and $\mathbf{a}$. 

The training procedure of $S_2$ is illustrated in Figure \ref{fig:main}(b). It adopts a self-supervised training paradigm, where the first three seconds of each audio sample is truncated as prompt, and the acoustic language model is trained to predict the acoustic units of the remaining part conditioned on its semantic units and the prompt acoustic units with cross-entropy loss. This in-context learning approach enables the model to grasp the correspondence in acoustic characteristics between the two parts and acquire style transfer ability. During inference, we use semantic tokens from the previous stage and acoustic units of source speech as the style prompt to realize cross-lingual style transfer.

\subsection{Unit-to-Wave Generation}

\begin{table*}[t]
    \centering
    \resizebox{\textwidth}{!}{
    \begin{tabular}{clccccc}
    \toprule
     ID & Model  & BLEU (Fr-En) ($\uparrow$) & BLEU (Es-En) ($\uparrow$) & SIM ($\uparrow$) & MOS ($\uparrow$) & SMOS($\uparrow$)    \\
    \midrule 
    1 & S2UT & 18.08 & 23.78 & / & 3.73 $\pm$ 0.05  & /  \\
    2 & S2UT + PPG-VC & 17.03 & 23.03 & 0.69 & 3.37 $\pm$ 0.07 & 3.30 $\pm$ 0.06   \\
    3 & S2UT + NANSY & 18.21 & 23.48 & 0.68 & 3.56 $\pm$ 0.06 & 3.47 $\pm$ 0.05   \\
    4 & S2UT + YourTTS & 16.23 & 21.09 & 0.69 & 3.74 $\pm$ 0.05 & 3.60 $\pm$ 0.06    \\
     \midrule
    5 &  Ours & 16.30 & 22.00 & \textbf{0.73} & 3.86 $\pm$ 0.06 & \textbf{3.69 $\pm$ 0.05}  \\
      \midrule
    6 &  Target Audio (CVSS-C) & 84.36 & 86.48 & / & 3.92 $\pm$ 0.05 & /  \\
    7 &  Target Audio (CVSS-T) & 80.99 & 82.12 & 0.69 & 3.95 $\pm$ 0.05 & 3.56 $\pm$ 0.06  \\
    \bottomrule 
    \end{tabular}
    }
    \caption{Results on translation quality and audio similarity on CVSS dataset.}
    \label{quality}
\end{table*}

\begin{table*}[t]
    \centering
    \begin{tabular}{clccc}
    \toprule
    ID & Model & SIM ($\uparrow$) & MOS ($\uparrow$) & SMOS ($\uparrow$)  \\
    \midrule
    1 & LibriTTS &  0.67 &  3.84 $\pm$ 0.05  & 3.55 $\pm$ 0.05   \\
    2 & Libri-Light unlab-60k & 0.73 &  3.86 $\pm$ 0.05 & 3.69 $\pm$ 0.05  \\
    3 & \ \ \ + CVSS source & 0.78 &  3.85 $\pm$ 0.05 & 3.74 $\pm$ 0.06  \\
    \bottomrule 
    \end{tabular}
    \caption{Ablation results on different compositions of training data.} 
    \label{datasize}
\end{table*}

In the waveform generation stage $S_3$, we adopt a GAN-based unit vocoder to map the target acoustic units to high-fidelity waveforms. Our vocoder is derived from BigVGAN \cite{lee2022bigvgan}, with a generator built from a set of look-up tables (LUT) that embed the discrete units, and a series of blocks composed of transposed convolution and a residual block with dilated layers. Multi-period discriminator (MPD) and multi-resolution discriminator (MRD) are used for adversarial training.

\section{Experiments}
\label{sec:exp}

\subsection{Setup}
\label{sec:setup}

\textbf{Datesets} We use two language pairs in the CVSS dataset \cite{jia2022cvss} as the translation benchmark, which are French-English (Fr-En) and Spanish-English (Es-En). For $S_2$ and $S_3$ stages, we use the \textit{unlab-60k} subset of Libri-Light \cite{kahn2020libri} to train the acoustic language model, and use LibriTTS \cite{zen2019libritts} to train the SoundStream model and the vocoder. All audio is processed at a 16 kHz sampling rate. We provide more details about the datasets in Appendix \ref{sec:data}.
    
\noindent \textbf{Model Configurations} We apply the publicly available multilingual HuBERT (mHuBERT) model\footnote{\url{https://dl.fbaipublicfiles.com/hubert/mhubert_base_vp_en_es_fr_it3.pt}} with the k-means model of 1000 clusters for the 11th-layer features\footnote{\url{https://dl.fbaipublicfiles.com/hubert/mhubert_base_vp_en_es_fr_it3_L11_km1000.bin}} and train a SoundStream model with a size of 1024 for each codebook and an overall downsampling rate of 320. For stage $S_1$, we train an S2UT-conformer for Fr-En following \cite{lee2021direct}, and follow the model in \citet{popuri2022enhanced} for Es-En but without mbart-decoder initialization. The decoder-only transformer of $S_2$ has about 760M parameters, with details of its configurations provided in Appendix \ref{sec:modelsetting}.

\noindent \textbf{Baselines} Considering that previous S2ST models with style transfer  \cite{jia2021translatotron, song2023styles2st, rubenstein2023audiopalm, barrault2023seamless} either differ from ours in settings or are not open-sourced, we mainly compare our model with S2UT models used in $S_1$ followed by a single-speaker vocoder\footnote{\url{https://github.com/facebookresearch/fairseq/blob/d9a627082fd03ec72a27a31a4e56289bfcb2e4e4/examples/speech_to_speech/docs/textless_s2st_real_data.md\#unit-based-hifi-gan-vocoder}, English version}, and cascaded pipelines formed by appending various voice conversion models after the vocoder, which are PPG-VC \cite{liu2021any}, NANSY \cite{choi2021neural} and YourTTS \cite{casanova2022yourtts}. 
    
\noindent \textbf{Evaluation Metrics} We employ both objective and subjective metrics to measure the model performance in terms of translation accuracy, speech quality, and style similarity with the source speech. For objective evaluation, we calculate the BLEU score between the ASR-transcripts of the translated speech and reference text as well as speaker cosine similarity (SIM). For subjective metrics, we use crowd-sourced human evaluation with 1-5 Likert scales and report mean opinion scores on speech quality (MOS) and style similarity (SMOS) with 95\% confidence intervals (CI). More details are provided in Appendix \ref{sec:eval}.

\subsection{Results and Analysis}

Table \ref{quality} summarizes the main experiment results. In terms of audio quality, our model achieves a high MOS of 3.86, surpassing baselines 2-4. This demonstrates the significant advantage of our model in speech naturalness compared to cascaded pipelines with voice conversion models. Moreover, our model gets higher MOS than direct S2UT, indicating that incorporating acoustic unit modeling helps improve the long-term naturalness of speech. On the other hand, our model achieves the highest speaker similarity, with SMOS being 3.69 and SIM being 0.73, which surpasses all three cascaded systems and even the CVSS-T target, demonstrating the outstanding performance in zero-shot cross-lingual style transfer of our model. This can be attributed to the large model size and extensive training data, through which our model acquires strong zero-shot style transfer capability and can generalize effectively to unseen source languages.

In terms of translation accuracy, generally, there is a comprehensive decrease in BLEU scores for 2-5 compared to 1, indicating that additional style transfer processes lead to a loss in semantic content. Compared to PPG-VC and NANSY, YourTTS and our model suffer from lower BLEU scores. We observe that this is due to the acoustic environment transfer capabilities of YourTTS and our S2 stage model, which transfer some of the strong background noise from the source speech into the generated speech, posing a challenge for ASR. Nevertheless, our model still maintains good translation accuracy, with BLEU declination restricted to 1.78 for both Fr-En and Es-En, outperforming the cascaded baseline with YourTTS.

\subsection{Ablation Studies}

We further conduct ablations on different training data compositions of $S_2$, and the results are summarized in Table \ref{datasize}. We observe that when using LibriTTS with a smaller size and fewer speakers, there is a significant decrease in SMOS and SIM of 0.14 and 0.06, with only a minor decrease in MOS of 0.02. This suggests that the model's style transfer performance relies on a large amount of speech data from multiple speakers while achieving high-quality speech generation does not require as much data. 

We also add part of the speech from the CVSS source to the training data to examine the model performance on unseen / seen speakers. We observe a gap of 0.05 for both SIM and SMOS. 
This indicates that our model's zero-shot style similarity still lags behind that of seen speakers. This gap can be narrowed by using a training corpus with more speakers.

\section{Conclusions}
\label{sec:conclu}

We propose an S2ST approach with style transfer capability by adopting an acoustic language model that learns style transfer through in-context learning. By adopting self-supervised training and large-scale training data, our method addresses the scarcity of speaker-parallel data and achieves cross-lingual style transfer with unseen source languages. Experiments indicate that our approach achieves outstanding results in terms of speech quality and style similarity while keeping good translation accuracy. 

\section{Limitations and Potential Risks}
\label{sec:limit}

Despite that our model excels in style transfer and generating high-quality translated speech, it still suffers from several limitations: 1) Our evaluation (especially the objective evaluation) of style transfer capability mainly focuses on the global speaker timbre, and we have not yet delved deeply into other stylistic characteristics such as prosody and emotion. We leave the exploration of these aspects for future work. 2) The large model size and the autoregressive generation paradigm may lead to efficiency issues, such as long inference latency. 3) The BLEU scores heavily depend on the ASR quality, which may not accurately reflect the speech translation performance. Future directions could be improving ASR quality or exploring other evaluation metrics without reliance on ASR models. Besides, due to the speaker timbre transfer capability of our model, it may be misused to disinform, defame, or commit fraud. We will add some constraints to guarantee people who use our code or pre-trained model will not use the model in illegal cases.

\section*{Acknowledgements}
This work is supported by National Key R\&D Program of China under Grant No.2022ZD0162000, National Natural Science Foundation of China under Grant No. 62222211 and No.62072397. 

\bibliography{custom}

\appendix
\section{Datasets}
\label{sec:data}

In this section, we provide details of the translation benchmark dataset and the corpora for training $S_2$ and $S_3$ models. 

\noindent \textbf{CVSS} CVSS \cite{jia2022cvss} is an S2ST benchmark dataset derived from the CoVoST 2~\citep{wang2020covost} speech-to-text translation corpus by synthesizing the translation text into speech using TTS systems. It comprises two sub-versions of CVSS-C and CVSS-T, where the target speech in CVSS-C is generated by a single-speaker TTS system while that of CVSS-T is generated by a multi-speaker TTS system with speaker timbre transferred from the source speech. We use CVSS-C for training and evaluating the translation models, and provide results of ground truth target audios in CVSS-T as a reference for style transfer performance.

\noindent \textbf{Libri-Light} Libri-Light is a large-scale corpus containing unlabelled speech from audiobooks in English. The \textit{unlab-60k} subset we use consists of 57.7k hours of audio with 7,439 speakers.

\noindent \textbf{LibriTTS} LibriTTS is a multi-speaker English TTS dataset. It comprises 585.5 hours of audio with 2,456 speakers.

\begin{figure}[t]
\includegraphics[width=\columnwidth]{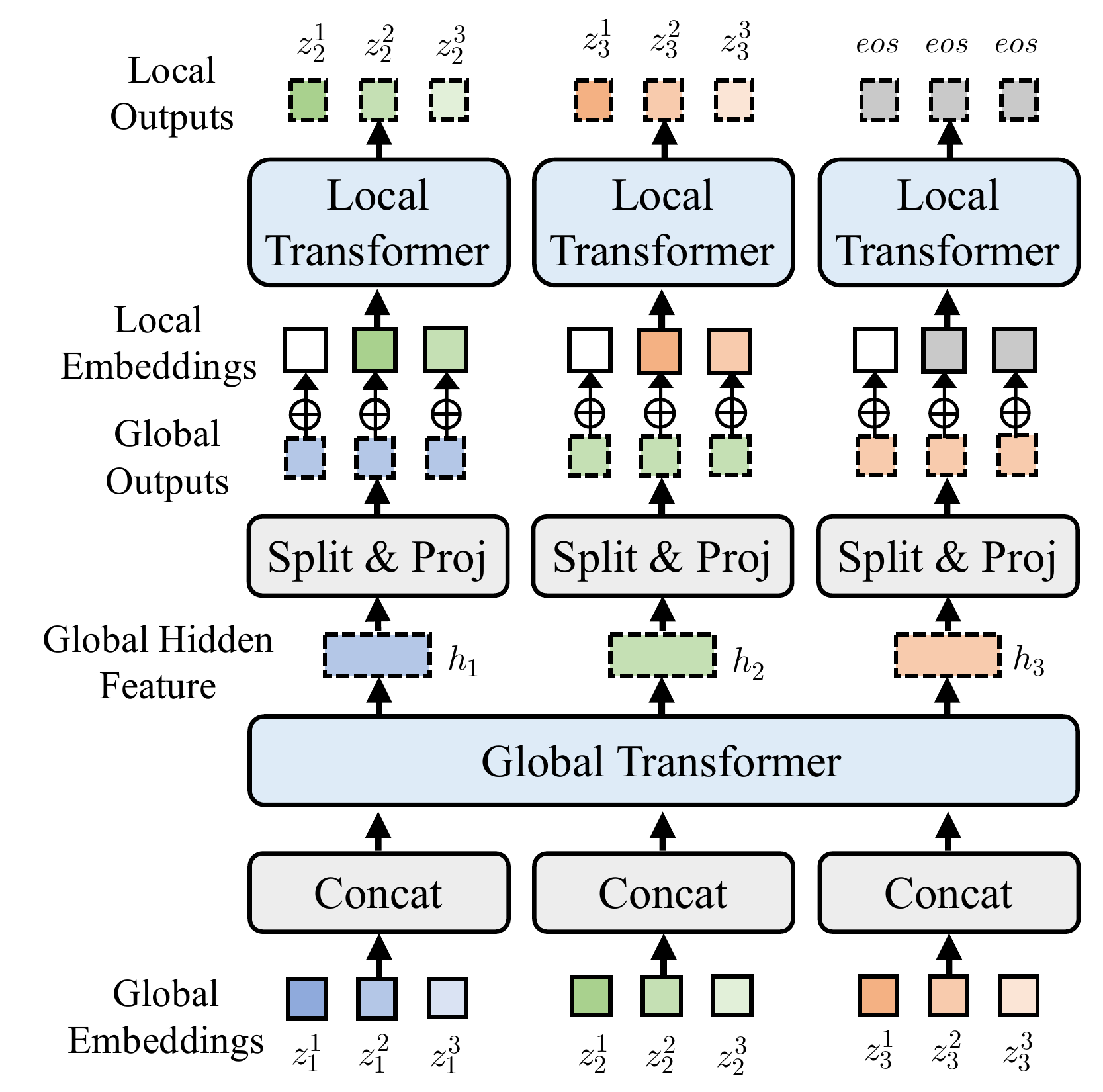}
\caption{The multi-scale architecture of UniAudio used for the $S_2$ stage model.}
\label{fig:multiscale}
\end{figure}

\begin{figure}[t]
\centering
\includegraphics[width=0.85\columnwidth]{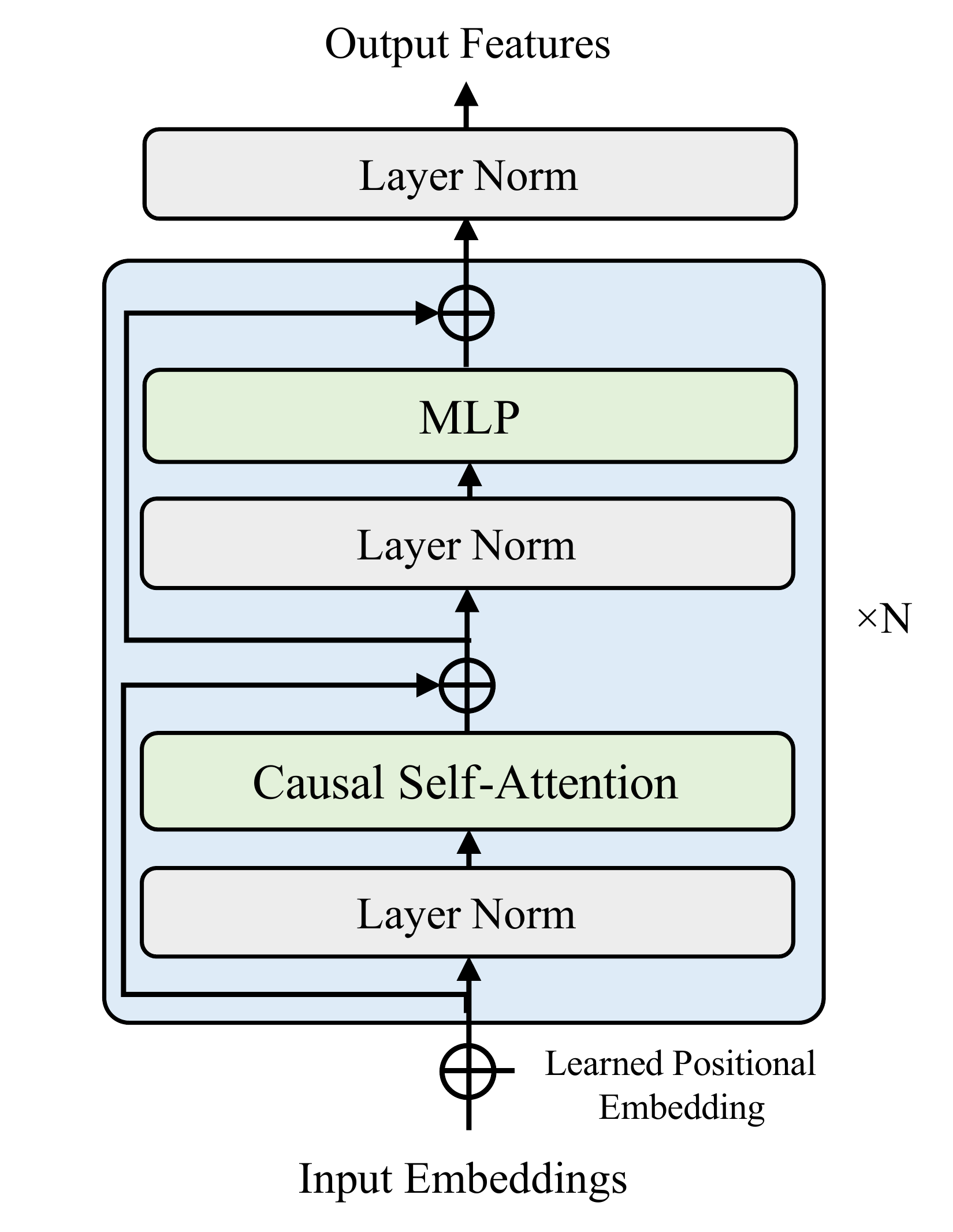}
\caption{Structure of the global transformer.}
\label{fig:global}
\end{figure}

\section{Model Settings}

\subsection{$S_2$ Model Architecture}
\label{sec:uniaudio}

UniAudio \cite{yang2023uniaudio} is a decoder-only transformer with an end-to-end differentiable multi-scale architecture to facilitate the modeling of long sequences. It has a hierarchical structure consisting of a global transformer and a local one. Figure~\ref{fig:multiscale} illustrates its multi-scale design. This model has exhibited remarkable capabilities in audio synthesis and modeling intrinsic relationships between acoustic and other modalities, as well as high efficiency in generating long sequences based on sub-quadratic self-attention. In this work, we adopt UniAudio as our $S_2$ stage model.

The architecture of the global transformer is illustrated in Figure \ref{fig:global}. The local transformer shares the same structure as the global one with two differences: 1) the local transformer has no positional embedding, and 2) there is a linear lm-head appended to the top for token prediction.

\subsection{Model Parameters}
\label{sec:modelsetting}

We provide hyperparameters of our $S_2$ and $S_3$ stage models in Table \ref{tab:hyperparameters}. We also refer the readers to the original papers \cite{lee2021direct, popuri2022enhanced} for details of $S_1$ models used. Each sub-module is trained with 4 NVIDIA-V100 GPUs for about a week.

\begin{figure*}[tb]
\begin{center}
    \includegraphics[width=\textwidth]{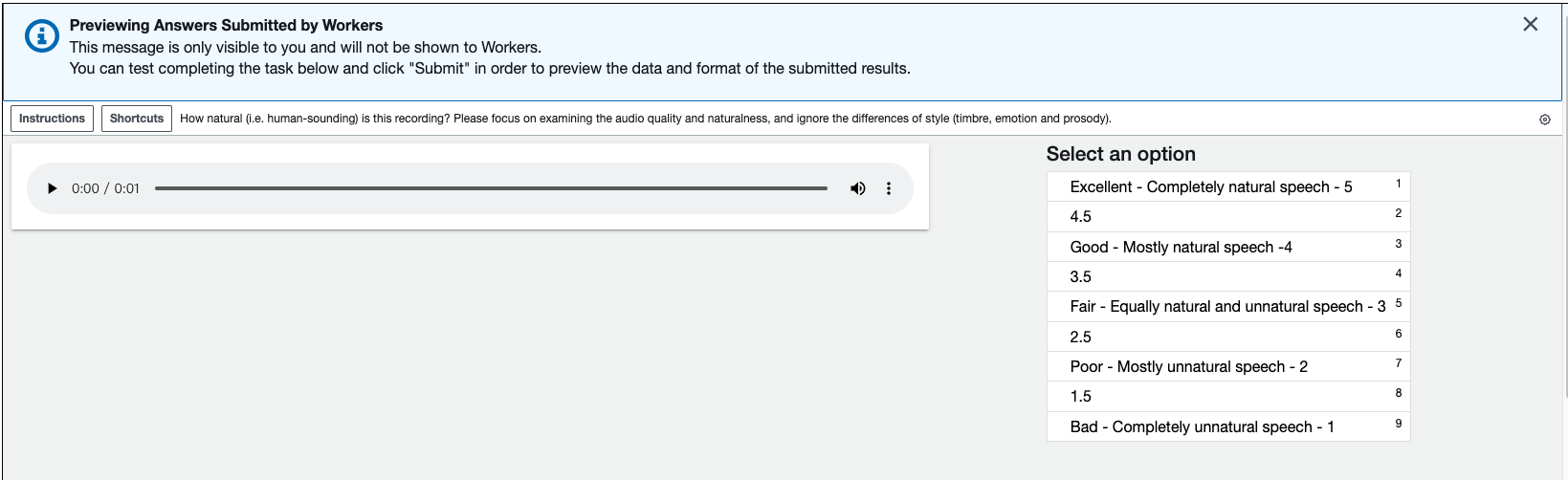}
\caption{Screenshot of MOS testing.}
\label{fig:mos}
\end{center}
\end{figure*}

\begin{figure*}[tb]
\begin{center}
    \includegraphics[width=\textwidth]{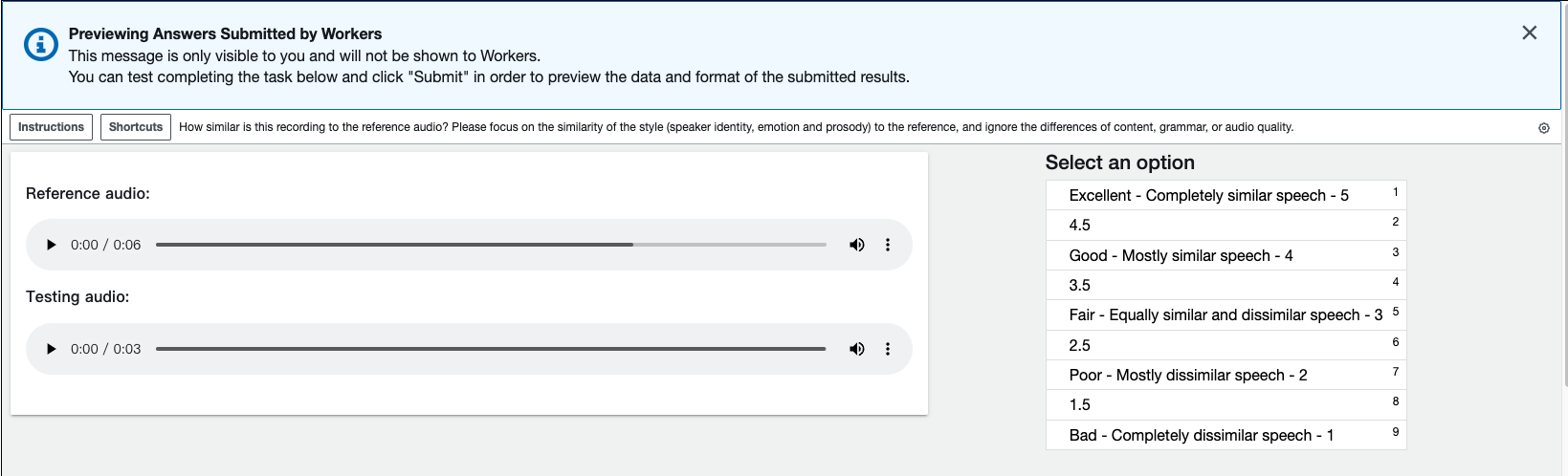}
\caption{Screenshot of SMOS testing.}
\label{fig:smos}
\end{center}
\end{figure*}

\section{Evaluation Metrics}
\label{sec:eval}
For translation accuracy, we use an open-sourced ASR model in \textit{fairseq} \footnote{\url{https://github.com/facebookresearch/fairseq/tree/main/examples/speech_to_speech/asr_bleu}} \cite{ott2019fairseq} to transcribe the audios and then calculate the BLEU score between the transcripts and the reference text. For speaker similarity, we use Resemblyzer\footnote{\url{https://github.com/resemble-ai/Resemblyzer}}, which is a public-available speaker encoder to extract speaker embeddings of the synthesized and source speech and calculate their cosine similarity.

Our subjective evaluation tests are crowd-sourced and conducted via Amazon Mechanical Turk. For audio quality evaluation, we ask the testers to examine the audio quality and naturalness. For style similarity, we instruct the testers to evaluate the style similarity between the synthesized and source speech while ignoring the content. The testers rate scores on 1-5 Likert scales. We provide screenshots of the testing interfaces in Figure \ref{fig:mos} and \ref{fig:smos}. Each data item is rated by 2 testers, and the testers are paid \$8 hourly.

Due to the large cost of conducting voice conversion and evaluation on the whole test split, we randomly sample 488 items from each language pair for evaluation, which represents approximately 3\% of the test set.

\begin{table}[ht]
    \small
    \centering
    \begin{tabular}{c|c|c}
    \toprule
    Model & \multicolumn{2}{c}{Hyperparameter}   \\ 
    \midrule
    \multirow{6}{*}{\tabincell{c}{Acoustic \\ Language \\ Model}} 
    &Global Layers         &    20     \\
    &Local Layers & 6 \\
    &Hidden Dim                   &   1,536  \\    
    &Attention Headers           &  16   \\  
    &FFN Dim              &  6,144   \\    
    &Number of Parameters               & 763.1M \\
    \midrule
    \multirow{4}{*}{\tabincell{c}{Unit\\Vocoder}} 
    &Upsample Rates                     &    [5,4,2,2,2,2]     \\
    &Hop Size                 &   320  \\    
    &Upsample Kernel Sizes          &  [9,8,4,4,4,4]   \\
    &Number of Parameters               & 121.6M \\

    \bottomrule
    \end{tabular}
    \vspace{2mm}
    \caption{Hyperparameters of $S_2$ and $S_3$ Stage Models.}
    \label{tab:hyperparameters}
\end{table}

\end{document}